\documentclass[12pt,preprint]{aastex}

\shorttitle{Evolving Starburst Models of Gas Media}
\shortauthors{Yao et al.}

\begin{document}

\title{Starburst Models For FIR/sub-mm/mm Line Emission. I. An Expanding Supershell Surrounding A Massive Star Cluster}

\author{Lihong Yao\altaffilmark{}}
\affil{Department of Astronomy and Astrophysics, University of Toronto,
    Toronto, ON M5S 3H8, Canada}
\email{yao@astro.utoronto.ca}

\author{T. A. Bell, S. Viti, J. A. Yates}
\affil{Department of Astronomy and Astrophysics, University College London,
    London WC1E 6BT, UK}

\and

\author{E. R. Seaquist}
\affil{Department of Astronomy and Astrophysics, University of Toronto,
    Toronto, ON M5S 3H8, Canada}

\author{}
\affil{Submitted 22 July 2005; Accepted by ApJ 19 September 2005}

\begin{abstract}
The effect of a newly born star cluster inside a giant molecular cloud (GMC) 
is to produce a hot bubble and a thin, dense shell of interstellar gas and dust 
swept up by the H II expansion, strong stellar winds, and repeated supernova 
explosions. Lying at the inner side of the shell is the photodissociation region
(PDR), the origin of much of the far-infrared/sub-millimeter/millimeter 
(FIR/sub-mm/mm) radiation from the interstellar medium (ISM). We present a model 
for the expanding shell at different stages of its expansion which predict 
mm/sub-mm and far-IR emission line intensities from a series of key molecular 
and atomic constituents in the shell. The kinematic properties of the swept-up 
shell predicted by our model are in very good agreement with the measurements 
of the supershell detected in the nearby starburst galaxy M 82. We compare the 
modeling results with the ratio-ratio plots of the FIR/sub-mm/mm line emission 
in the central 1.0 kpc region to investigate the mechanism of star forming activity 
in M 82. Our model has yielded appropriate gas densities, temperatures, and 
structure scales compared to those measured in M 82, and the total H$_2$ content 
is compatible with the observations. This implies that the neutral ISM of the 
central star-forming region is a product of fragments of the evolving shells.
\end{abstract}


\section{Introduction} \label{intro}

Starburst is a phenomenon when the star formation rate (SFR) cannot be sustained 
for the lifetime of the galaxy. It is now clear that active star formation or 
starburst activity is common throughout the universe \citep{mad98}. The bursts 
of massive star formation can dramatically alter the structure and evolution of 
their host galaxies by injecting large amounts of energy and mass into the ISM 
via strong stellar winds and repeated supernova explosions. The evolution of 
the superbubbles and supershells that have sizes ranging from several tens to 
hundreds of parsec plays an important role in understanding the amount and 
distribution of warm gas in the ISM. Furthermore, understanding the characteristics 
of starbursts and their relationship with the ISM, as well as to be able to 
parametrize the star formation history are crucial in understanding the 
galaxy evolution. 

In the past, several models have been used to interpret the infrared, sub-millimeter, 
millimeter line observations of neutral gas in the central regions of nearby 
starburst galaxies \citep[e.g.][and references therein]{mao00, saf00, wil92}. 
These include the large velocity gradient (LVG) model \citep{gak74}, the 
steady-state PDR model \citep{tah85}, and the inhomogeneous radiative transfer 
model taking into account non local thermodynamic equilibrium (non-LTE) \citep{wil92}. 
These have revealed that the physical conditions (such as gas density, FUV flux, 
and gas kinetic temperature) are enhanced in starburst regions. However, none of 
these models are able to relate the observed line emission properties of the 
neutral gas in a starburst galaxy to its age and star formation history. 
In this paper, we introduce an evolving starburst model for FIR/sub-mm/mm line 
emission in gas media that allows us to ultimately achieve this goal.

\begin{figure}
\epsscale{1.0}
\plotone{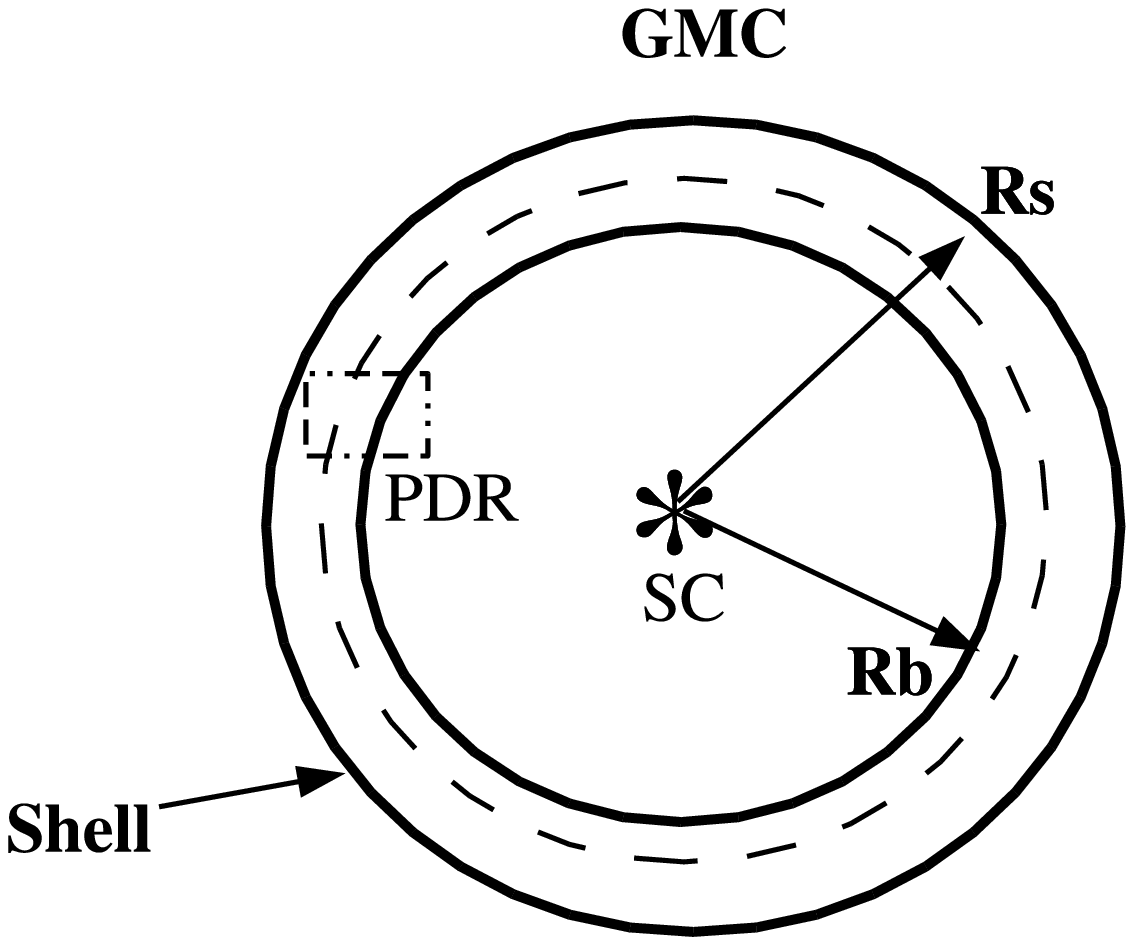}
\caption{A schematic structure of an evolving GMC centrally illuminated by a compact 
young star cluster (SC). $R_s$ is the radius of the shell, and $R_b$ is the radius 
of the bubble. The PDR lies between the thin, dense swept-up shell and the interior
\citep{hat97}. 
\label{shell}}
\end{figure}

Our model consists of a standard dynamical model of the bubble/shell structure 
around a young star cluster (see Fig.~\ref{shell}), which has been described 
in many publications \citep[e.g.][]{cmw75, wea77, mak87, fra90, kam92}, 
a time-dependent stellar population synthesis model \citep{lei99}, 
a fully time-dependent PDR chemistry model \citep{bel05}, and a one-dimensional 
non-LTE line radiative transfer model \citep{ray01}. In this paper, we conduct 
a preliminary study using this set of models. We first describe the methodology 
of our model (Section 2). We then follow the evolution of a GMC and a swept-up 
shell induced by massive star formation at the center, and calculate the dynamics, 
thermal structure, and the line radiative transfer of the selected molecular 
and atomic species in the expanding shell (Section 3). We compare our modeling 
results with the observations of the expanding supershell and average gas 
properties in the central 1.0 kpc region of the nearby starburst galaxy M 82 
(Section 4). Finally, we present the conclusions of this study (Section 5). 

The basic assumptions for our evolving starburst model are (1) star formation 
occurs primarily within the dense optically thick spherical cloud 
\citep[e.g.][]{gao01}, and that all stars form instantaneously in a compact 
spherical cluster located at the center of the cloud (the star cluster is 
therefore treated as a point source), and (2) the starlight produced by the 
central cluster is completely absorbed and reprocessed by the dust in the 
expanding shell \citep{efs00}. A summary of our evolving starburst model is 
presented in Table~\ref{tbl-1}. 

\section{Starburst Models For Gas Media} \label{model}

The evolution of a giant molecular cloud is determined by H II expansion in 
the very early stage ($t$ $<$ 10$^5$ yr), when a hot bubble surrounded by a 
thin dense shell structure is created. The later evolution is driven by the 
strong stellar winds and repeated supernova explosions. We assume that repeated 
supernova explosions behave like a steady isotropic stellar wind injected to 
the bubble. The hot bubble will eventually cool, and the swept-up shell will 
stall after a few times 10$^7$ yr. The stars in the young cluster located at 
the center of the GMC are assumed to have masses between 0.1 M$_{\odot}$ and 
120 M$_{\odot}$. The Salpeter initial mass function 
$dN/dm_{\ast}$ $\propto$ $m^{-2.35}_{\ast}$ \citep[IMF;][]{sal55} is adopted
in this study. A top-heavy IMF, which has an excess of stars in the mass range 
10 - 20 M$_{\odot}$ over stars of 5 M$_{\odot}$ or less for starburst galaxies
\citep[e.g.][]{rie80}, will be investigated in future work.

\subsection{Shell Dynamics} \label{dyn}

The radius and velocity of the {\it H II Expansion} due to ionization can be 
written as \citep{spi78, fra90},
\begin{mathletters}
\begin{eqnarray}
R_{HII}(t) & = & R_S\Big(1 + \frac{7}{4}\frac{c_i t}{R_S}\Big)^{\frac{4}{7}}, \\
V_{HII}(t) & = & c_i \Big(1 + \frac{7}{4}\frac{c_i t}{R_S}\Big)^{-\frac{3}{7}}
\end{eqnarray}
\end{mathletters}

where $R_S$ is the initial Str\"{o}mgren radius in pc, and $c_i$ $\simeq$ 11.5 
km s$^{-1}$ is the sound speed in the ionized gas with an equilibrium temperature 
of $\sim$ 10$^4$ K. 

Almost as soon as the initial Str\"{o}mgren sphere is formed, the strong 
winds start to impart large amounts of mechanical energy into the bubble. 
About 96\% of the total wind energy is generated by stars with masses $>$ 30 
$M_\odot$ \citep{mak87}. The size of the hot bubble is assumed to be much 
larger than the thickness of the swept-up shell, therefore the radius and 
velocity of the shell in the {\it Winds} phase can be written as \citep{mak87}, 
\begin{mathletters}
\begin{eqnarray}
R_w(t) & = & 269.0 \Big(\frac{L_{38}}{n}\Big)^{\frac{1}{5}} (t_7)^{\frac{3}{5}}, \\ 
V_w(t) & = & 16.1 \Big(\frac{L_{38}}{n}\Big)^{\frac{1}{5}} (t_7)^{-\frac{2}{5}} 
\end{eqnarray}
\end{mathletters}

where $L_{38}$ = $L_w$/(10$^{38}$ ergs s$^{-1})$, $L_w$ is the wind mechanical 
luminosity, $L_w$ = $\int_{m_{1}}^{m_{2}} C_w C_m m^{\gamma - 2.35}_{\ast} 
dm_{\ast}$, $t_7$ = $t$/(10$^7$ yr), $n$ is the ambient gas density in cm$^{-3}$,
$m_{1}$ = 0.1 M$_{\odot}$, $m_{2}$ = 120 M$_{\odot}$, $C_w$ = 1.0 $\times$ 10$^{29}$, 
$C_m$ = 429.0, and $\gamma$ = 3.7 (derived from Abbott 1982). The main-sequence 
lifetime of the most massive star (120 M$_{\odot}$) in the star cluster is about 
7.0 $\times$ 10$^5$ yr \citep{mam88}. After this time, we assume that the 
winds-equivalent energy produced by the first supernova and the subsequent ones 
drives the further expansion of the swept-up shell. The radius and velocity of 
the shell in the {\it Supernova} phase can be written as,  
\begin{mathletters}
\begin{eqnarray}
R_{SN}(t) & = & 97.0 \Big(\frac{N_{\ast} E_{51}}{n}\Big)^{\frac{1}{5}} \Big[(t_7)^{\frac{3}{5}} - \Big(\frac{t_{1stSN}}{10^7}\Big)^{\frac{3}{5}}\Big] + R_w(t_{1stSN}), \\
V_{SN}(t) & = & 5.7 \Big(\frac{N_{\ast} E_{51}}{n}\Big)^{\frac{1}{5}} (t_7)^{-\frac{2}{5}}
\end{eqnarray}
\end{mathletters}

where $N_{\ast}$ is the number of stars with masses $\ge$ 8 M$_{\odot}$ in the 
cluster, $E_{51} = E_{SN}$/(10$^{51}$ ergs s$^{-1})$, $E_{SN}$ is the energy 
produced by each supernova explosion, $t_{1stSN}$ is the time when the first 
supernova occurs in the star cluster, and $R_w$($t_{1stSN}$) is the shell radius 
at $t_{1stSN}$ calculated from Equation (2a). The average rate of supernova explosions 
is $\sim$ 6.3 $\times$ 10$^{35}$ $N_{\ast} E_{51}$ ergs s$^{-1}$ \citep{mak87}. 
When the energy produced by the stellar winds and/or supernova explosions is much 
greater than the radiative losses, the bubble is adiabatic. This {\it Adiabatic} 
phase persists until the radiative cooling becomes important for the hot bubble 
at $t_c$,

\begin{equation}
t_{c} = 4 \times 10^6 \mathcal{Z}^{-1.5} (N_{\ast} E_{51})^\frac{3}{10} n^{-\frac{7}{10}} 
\end{equation}

where $\mathcal{Z}$ is the metallicity with respect to the solar. After $t_c$, 
the expansion of the bubble is no longer energy-driven, but momentum-driven. 
This momentum-driven phase is characterized as the snow-plow (SP) phase. 
For simplicity, we ignore the momentum deposition in the shell by SN ejecta 
\citep{mak87}. Hence, the radius and velocity of the shell in the {\it Snow-plow} 
phase can be written as,
\begin{mathletters}
\begin{eqnarray}
R_{SP}(t) & = & R_{c} \Big(\frac{t}{t_{c}}\Big)^\frac{1}{4}, \\
V_{SP}(t) & = & \frac{R_{c}}{4 t_{c}} \Big(\frac{t}{t_{c}}\Big)^{-\frac{3}{4}}
\end{eqnarray}
\end{mathletters}

where $R_{c}$ is the radius of the bubble at cooling time $t_{c}$. The snow-plow 
phase ends when the shell expansion velocity is close to the thermal sound 
speed of gas in the ISM (typically $\sim$ 10 km s$^{-1}$). The shell will stall 
and disperse due to the Rayleigh-Taylor instability \citep{mac99}. 

Our one-dimensional shell dynamical model may overestimate the winds and 
supernova mechanical luminosities as argued recently by Dopita et al. (2005), 
because the mixing and dynamical instabilities will occur in two dimensions,
and the ISM is intrinsically inhomogeneous. Dopita et al. (2005) also suggested
that the higher ISM pressure in starburst regions causes the expanding shell to
stall at a smaller radius. Another argument is that the gravitational instability 
may induce new star formation inside the shells. These concerns may indicate 
that the conventional bubble/shell dynamics \citep{wea77, mak87} may need 
to be modified. 

\subsection{Physical Conditions of The Swept-up Gas} \label{cond}

The PDRs that lie at the inner sides of the clouds or shells centrally illuminated 
by massive star formation are the origin of much of the FIR/sub-mm/mm radiation 
from the ISM. Physical conditions of the swept-up gas in these PDRs are very 
different from those of the cold gas components in the ISM. The gas temperature 
and density of the swept-up shells are a few orders of magnitude higher due to 
the strong FUV radiation and shock compression. The FUV radiation (6 eV $<$ $h\nu$ 
$<$ 13.6 eV) produced by newborn stars plays an important role in the heating 
and chemistry of PDRs, especially during the early evolution. Other sources 
that may contribute to the shell heating are the mechanical energy input by 
winds and SN explosions \citep{mck99}, shocks caused by the accretion 
of gas at the outer surface of the shell \citep{mah80}, and cosmic rays 
\citep{suc93, bra03}. Cosmic rays may play an important role in the heating 
of swept-up gas after the stars with masses $\ge$ 8 M$_{\odot}$ have terminated 
as supernovae. Heating sources due to cloud-cloud collisions \citep{mns79} or 
shell-shell interaction \citep{saj99} are not considered in this study.  

The total FUV flux is calculated by integrating the flux of the stellar 
population spectrum between 912 \AA $ $ and 2055 \AA $ $ for each time step
using Starburst99, a time-dependent stellar population synthesis model developed
by Leitherer et al. (1999). We consider an instantaneous burst for the star
formation law, where the star formation occurs all at the same time (i.e. at 
age zero). The FUV field strength $G_0$ incident on the inner surface of the shell 
(visual extinction $A_v$ = 0) is then calculated by taking the ratio of the total 
FUV flux to the surface area 4$\pi$$r^2_s(t)$ of the expanding shell at each 
time-step. We use the same input parameters and assumptions for Starburst99 
as those used in the shell dynamics calculation (see Table~\ref{tbl-1}). 

The swept-up shell itself is supported by thermal gas pressure and non-thermal 
pressure due to micro-turbulence. The gas temperature decreases toward the
outer surface of the shell, and the total gas density is assumed uniform. Therefore 
the pressure is lower at the outer surface. The shell density $n_s$ refers to 
the total H$_2$ density $n$(H$_2$) in this study. The shell density at each 
time-step is derived from balancing the pressure at the outer surface of the 
shell with the ram pressure,

\begin{equation}
n_{s}(t) = \frac{n_a v^2_s(t)}{kT_{s}(t)/\mu + \delta v^2_D}
\end{equation}

where $n_a$ is the ambient molecular gas density, $v_s$($t$) is the expansion 
velocity, $T_s$($t$) is the gas temperature at the outer surface of the shell, 
$\mu$ is the mean molecular weight, $\mu$ = 0.62 $m_H$, $m_H$ is the mass of the 
hydrogen atom, and $\delta$$v_D$ is the micro-turbulent velocity inside the shell. 
The calculation of the gas temperature profile across the shell will be described 
in the following section. The thickness of the shell $d_s$ at each time-step is 
in turn calculated using the continuity equation (or mass conservation law),

\begin{equation} 
d_{s}(t) = \frac{n_a r_{s}(t)}{3 n_{s}(t)}
\end{equation}

\subsection{The Time-dependent PDR Model} \label{pdr}

The gas temperature and chemical abundances of the swept-up shell are calculated 
self-consistently at each depth- and time-step using the time-dependent PDR model 
developed at UCL (called UCL\_PDR). A fully time-dependent treatment of the chemistry 
is employed in UCL\_PDR which includes 128 species involved in a network of over 
1700 reactions \citep[][and references therein]{bel05}. The polycylic aromatic 
hydrocarbons (PAHs) chemistry is not included. The reaction rates are taken 
from the UMIST chemical database \citep{let00}. Detailed chemical modeling, 
heating and cooling mechanisms and the thermal balance between them are described 
in the literature \citep[e.g.][and references therein]{tay93, pap02}. Heating due 
to shocks is not included. The UCL\_PDR code has been modified for the purpose 
of this study to include a pressure balance check at the outer surface of the shell, 
as well as the evolution information of shell density, thickness, and FUV 
radiation strength. 

The UCL\_PDR code assumes a plane-parallel geometry and models the PDR as a 
semi-infinite slab of homogeneous density at a given time-step. The pressure
is thus not in equilibrium across the PDR region. The FUV radiation field 
illuminates the shell from one side, and it becomes attenuated with increasing 
visual extinction $A_v$ into the shell at a given time-step as 
$G$ = $G_0$$e^{-2.4 k A_v}$, where $G_0$ is the FUV strength at $A_v$ = 0 
calculated by the Starburst99 model. The coefficient $2.4 k$ in front of the 
$A_v$ in the exponent takes into account the difference in opacity from the 
visible to the UV and the influence of grain scattering. The timescale 
for gas in PDRs to reach chemical equilibrium depends on the gas density, 
temperature, degree of ionization, and species involved \citep{hat97, vab98}. 
In our study, this timescale varies from 10$^5$ to 10$^7$ yr for the 
swept-up gas. Our comparative tests using a single time-step model  
fail to reproduce important chemical structure features predicted by the 
fully time-dependent model for ages up to 10 Myr. The use of a full 
time-dependent PDR code in which temperature and density changes with time
is therefore justified in modeling the shell evolution over these timescales.

\subsection{The Non-LTE Line Radiative Transfer Model} \label{lrt}

The line radiative transfer properties are calculated using the Spherical Multi-Mol 
(SMMOL) code. The SMMOL model was also developed at UCL, implementing an accelerated 
$\Lambda$-iteration (ALI) method to solve multi-level non-LTE radiative 
transfer problems of gas inflow and outflow. The code computes the total radiation 
field and the level populations self-consistently. At each radial point, SMMOL
generates the level populations and the line source functions. Our model assumes 
that the gas emission originates in the unresolved, homogeneous, spherical expanding 
shell and that all gas and dust in the H II region have been swept up into the shell, 
i.e. a dustless H II region. The background radiation field is assumed to be the 
cosmic background continuum of 2.73 K. A detailed description of the SMMOL radiative 
transfer model and its implementation can be found in the appendix of 
Rawlings \& Yates (2001). The benchmarking and comparison with other 
line radiative transfer models are presented in van Zadelhoff et al. (2002). 
 
Several programs were developed to separate and extract the gas temperature 
and fractional abundances for molecular and atomic species calculated by the 
UCL\_PDR code. These extracted gas temperature and abundances, along with the 
shell density, thickness, radius, and expansion velocity computed by the 
dynamical code, are re-gridded for a spherical geometry and used as input 
parameters for the SMMOL code to compute the total line intensity or flux. 
Einstein A and collisional rate coefficients for the molecular and atomic 
lines are taken from the Leiden Atomic and Molecular Database \citep{sch05}. 
The lowest 10 energy levels are calculated for all molecules, 3 for atomic 
[C I] and [O I], and 8 for atomic [C II].

\section{Simulation of An Expanding Shell} \label{sim}

Observational studies have shown that molecular clouds in the Milky Way 
have a distinct mass spectrum $M^{\alpha}_{GMC}$, with $\alpha$ = -1.5 $\pm$ 0.1 
\citep{san85, sol87} for cloud masses ranging between 10$^2$ and 10$^7$ M$_{\odot}$. 
Therefore, about 70\% of the molecular mass in the Galaxy is contained 
in the GMCs with masses $>$ 10$^{6}$ M$_{\odot}$. These giant molecular clouds 
are known to be associated with active formation of massive stars. If we assume
that the cloud mass distribution in a starburst galaxy follows a similar index 
to the galactic one, we would expect much of the luminosity of the starburst 
to arise from the GMCs with a fairly narrow range of masses. We adopt a value 
for the cloud mass of 10$^7$ M$_{\odot}$ for the GMCs in this study. 
We assume the average gas consumption rate or star-formation efficiency 
$\eta$ in starburst galaxies per 10$^8$ yr to be 0.25 \citep{ken98}. Therefore, 
the total stellar mass $M_{\ast}$ for the star cluster in the center of the GMC 
is 2.5 $\times$ 10$^6$ M$_{\odot}$, and the number of stars $N_{\ast}$ with masses 
$m_{\ast}$ $\ge$ 8 M$_{\odot}$ is about 2.2 $\times$ 10$^4$. The radius of the 
GMC is about 50 pc with an average cloud density $n_0$ = 300 cm$^{-3}$ and 
a cloud core density $n_c$ = 2 $\times$ 10$^3$ cm$^{-3}$ \citep[][]{plu92, efs00}. 
The ambient density $n_{ism}$ is assumed to be 30 cm$^{-3}$ \citep{cat94}. 
Here we present an idealized case study with this particular set of input parameters.

\subsection{Kinematics of The Swept-up Gas} \label{kin}

The size of the H II region increases slowly with time. The Str\"{o}mgren 
radius is about 4.9 pc assuming the number of Lyman continuum photons is 
5 $\times$ 10$^{52}$ s$^{-1}$. The wind bubble catches up with the H II ionization 
front in a time less than 10$^5$ yr. The strong stellar winds cause the bubble 
to expand quickly into the cloud and sweep up more gas into the shell. 
The total wind power is estimated as $L_w$ $\simeq$ 1.4 $\times$ 10$^{40}$ 
erg s$^{-1}$ for the star cluster used in the model. When the most massive star 
in the center cluster (120 M$_{\odot}$) terminates as a supernova at $\sim$ 0.7 Myr, 
the thin shell caused by the H II region expansion and the stellar winds is
still expanding at a speed of $\sim$ 40 km s$^{-1}$. At this time, the shell 
has swept up much of the mass of its parent cloud, and is propelled into the ISM 
with a uniform density. The mechanical energy produced by the first supernova 
and the subsequent ones re-energizes the shell. A supernova cut-off mass of 
8 M$_{\odot}$ is assumed. The total energy generated by supernova explosions
is $\sim$ 2.0 $\times$ 10$^{55}$ ergs over 40 Myr. At $\sim$ 7.5 Myr, the 
hot bubble starts to cool and loses its internal pressure, at which time the 
adiabatic phase ends. We adopt 1.0 for the metallicity $\mathcal{Z}$ with respect 
to the solar throughout this study. The effect of lower metallicity, which is 
suspected to be present in starburst galaxies, will be discussed in a future paper. 
The radius and velocity of the shell at the end of the adiabatic phase are about 
270 pc and 24 km s$^{-1}$, respectively. At $\sim$ 50 Myr, the expansion velocity 
of the shell decreases to $\sim$ 10 km s$^{-1}$, the shell stalls and becomes
thicker and less dense.   

\begin{figure}
\epsscale{1.0}
\plotone{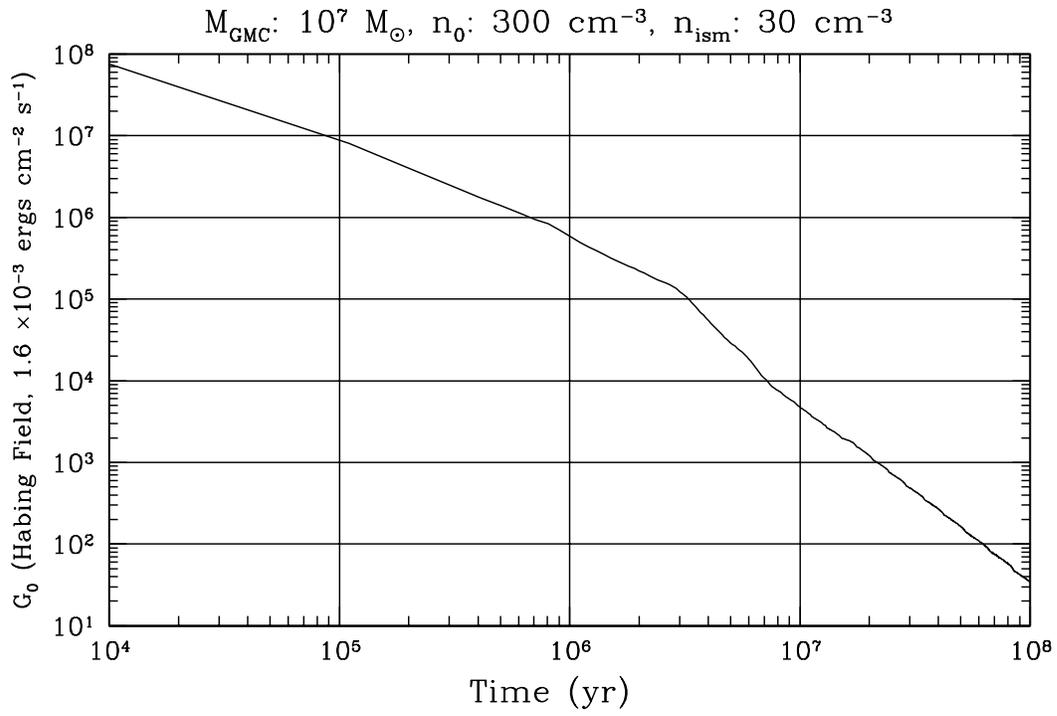}
\caption{Plot of the FUV radiation field strength $G_0$ incident on the inner surface 
of the shell ($A_v$ = 0) as a function of time (see text for details). \label{G0}}
\end{figure}

Fig.~\ref{G0} shows the FUV radiation strength $G_0$ incident on the inner surface 
of the shell ($A_v$ = 0) as a function of time. The $G_0$ value is in units of the 
Habing field, that is 1.6 $\times$ 10$^{-3}$ ergs cm$^{-2}$-s$^{-1}$ throughout
this study. The FUV strength decreases from about 10$^8$ to 10$^5$ from the onset 
of star formation to about 5 Myr when most of the massive O stars ($>$ 30 M$_{\odot}$) 
have terminated as supernovae. It then decreases twice as fast to a value of 40 
at 100 Myr.

PDRs are the origin of much of the FIR/sub-mm/mm line emission in a starburst 
galaxy. The surface layer ($A_v$ $\sim$ 1) contains atomic H, C, C$^{+}$ and O;
the transition from atomic to molecular hydrogen occurs at the center layer 
($A_v$ $\sim$ 1 - 2), whilst C$^{+}$ is converted into C and then CO over the 
region $A_v$ $\sim$ 2 - 4. H$_2$ and CO then extend to higher $A_v$ region and 
for $A_v$ $>$ 10 atomic O begins to be transformed into molecular O$_2$. 
The H$_2$ molecule provides effective self-shielding from the FUV radiation field. 
The CO layer also shows a degree of self-shielding, and therefore extends deeper 
into the shell. Small grains play an important role in the photoelectric heating 
of PDRs. Gas heating is dominated by collisional deexcitation of FUV-pumped
H$_2$ and vibrationally excited H$_2$ at the PDR surface. The thermal energy 
radiated by the dust is important for the gas heating at larger optical depth 
\citep{hol91}. The gas heating/chemistry at later evolutionary stages is no 
longer dominated by stellar radiation but by other sources, such as cosmic-rays 
and X-rays. The PDR cooling is dominated by fine structure line emission, such 
the [C II] 158$\mu$m and [O I] 63 $\mu$m transitions, whose critical densities 
are 3 $\times$ 10$^3$ cm$^{-3}$ and 5 $\times$ 10$^5$ cm$^{-3}$, respectively. 
At greater depths, molecular line emission (CO, OH, H$_2$O), ro-vibrational 
transitions of H$_2$, and gas-dust collisions contribute to the PDR cooling. 

\begin{figure}
\epsscale{1.0}
\plotone{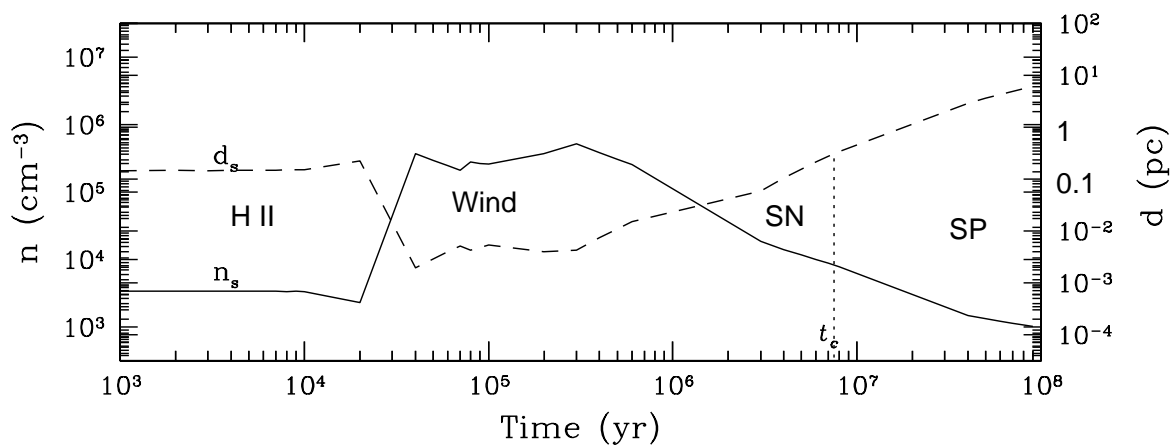}
\caption{Plot of the shell density ($n_s$ or $n$(H$_2$), solid line) and thickness 
($d_s$, dashed line) as a function of time for $M_{GMC}$ = 10$^7$ M$_{\odot}$, an 
initial cloud density $n_{GMC}$ = 300 cm$^{-3}$, and an ambient ISM density 
$n_{ism}$ = 30 cm$^{-3}$. The radiative cooling of the hot interior occurs at 
$t_c$ $\simeq$ 7.5 Myr (dotted line). Data for $n_s$ and $d_s$ after 10$^4$ years 
shown in the plots have been smoothed. 
\label{nsds}}
\end{figure}

\begin{figure}
\epsscale{1.0}
\plotone{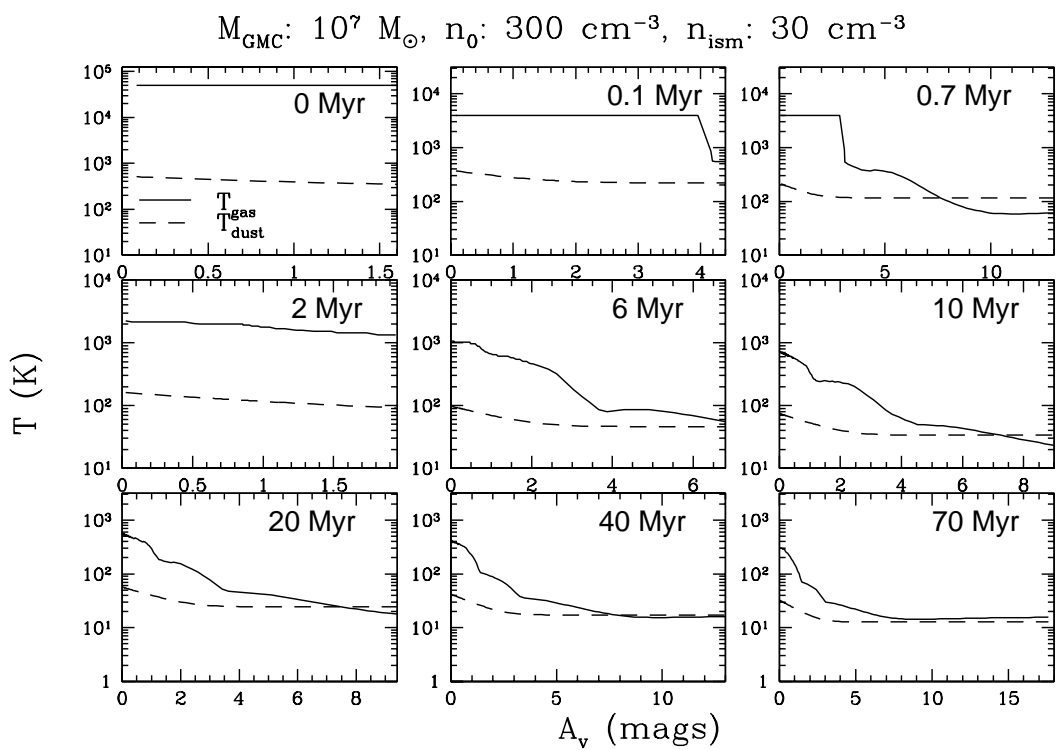}
\caption{The time-dependent gas and dust temperatures as a function of visual 
extinction $A_v$ for an expanding shell. Solid lines represent gas temperature, 
and dashed lines indicate dust temperature. \label{T2Av}}
\end{figure}

\begin{figure}
\epsscale{1.0}
\plotone{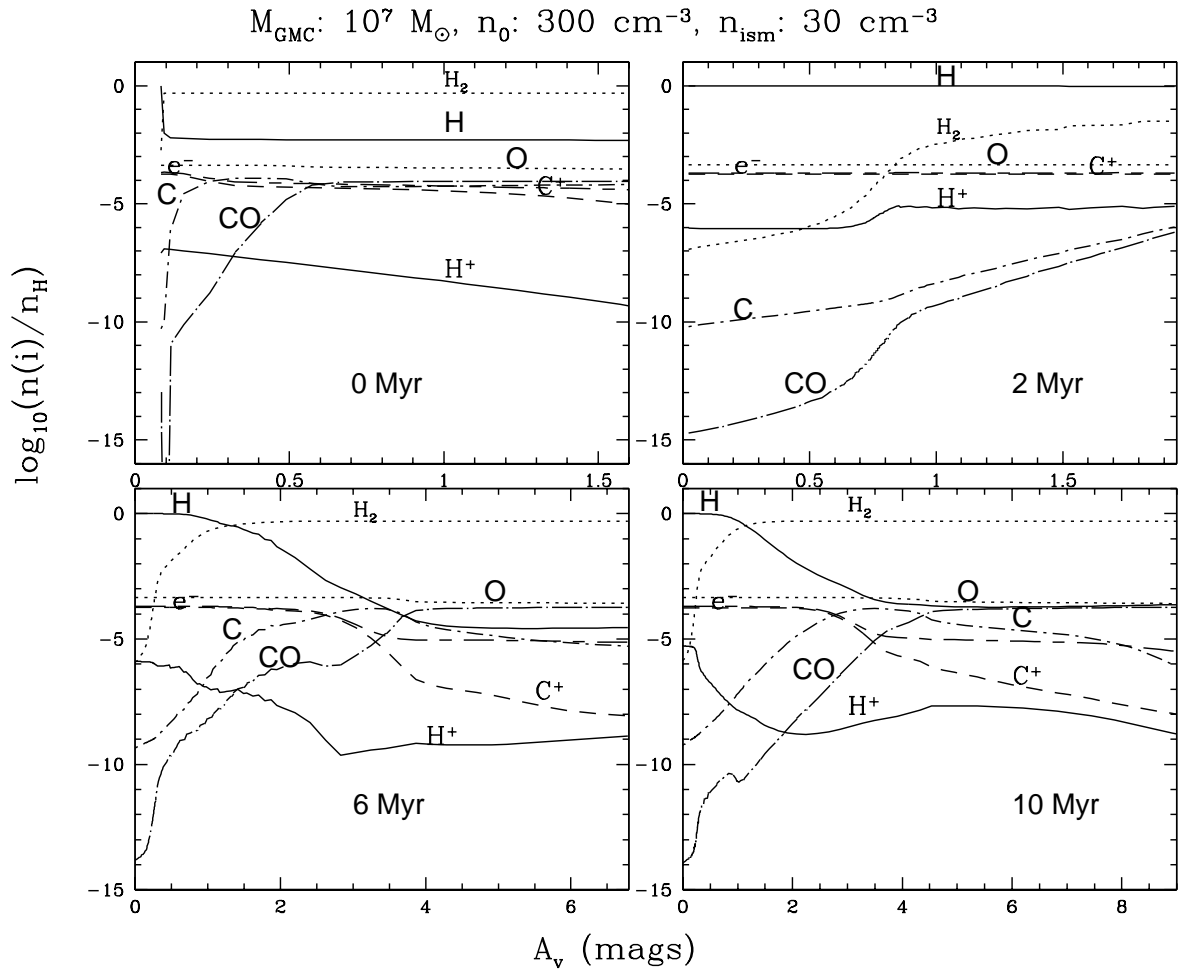}
\caption{The time-dependent chemical abundances of the main species (H, H$_2$, H$^+$, 
e$^-$, C, C$^+$, O, and CO) relative to the total hydrogen density, as a function of 
visual extinction $A_v$ for an expanding shell. \label{x2Av}}
\end{figure}

Table~\ref{tbl-2} summarizes the input parameters for our fully time-dependent 
PDR model. The initial abundance of H$_2$ is set to $n$(H$_2$)/$n_H$ = 0.5 
\citep{har03}. At the first time-step ($t$ = 0 yr) all depth-steps take as 
their initial abundances the values produced by a single-point dense dark-cloud
model. The input parameters for the dark-cloud modeling are $n_H$ = 4 $\times$ 10$^5$ 
cm$^{-3}$, $T_{GMC}$ = 10 K, $G_0$ = 1, and the gas-phase abundances relative 
to H nuclei $x_{He}$ = 7.5 $\times$ 10$^{-2}$, $x_C$ = 1.8 $\times$ 10$^{-4}$, 
$x_O$ = 4.4 $\times$ 10$^{-4}$, and $x_{Mg}$ = 5.1 $\times$ 10$^{-6}$. 
For subsequent time-steps, the input abundances are re-set to the output 
abundances of the previous time-step generated by the UCL\_PDR code. The gas 
temperature and chemical abundances at each depth- and time-step are calculated 
by balancing the heating and cooling. The cosmic-rays ionization rate is 
enhanced by a factor of 1.5 at later times ($t$ $>$ 10 Myr) to artificially 
include the soft X-rays heating effect on the gas of the shell. We assume that the 
gas-to-dust mass ratio is 100. Fig.~\ref{nsds} shows the shell density $n_s$ 
(or $n$(H$_2$)) and thickness $d_s$ as a function of time, as calculated 
by the shell dynamical code and the UCL\_PDR code, under the condition that the 
gas pressure at the outer surface of the shell differs from the ambient gas pressure 
by $\le$ 10\%. The shell density varies between 10$^3$ and 10$^6$ cm$^{-3}$, 
and the thickness of the shell changes from 10$^{-3}$ to 10 pc over the 100 Myr. 
We adopt a fixed micro-turbulent velocity $\delta$$v_D$ = 1.5 km s$^{-1}$ 
for the shell. The evolution of the shell density and thickness is constrained 
by the expansion velocity $v_s$, the shell temperature $T_s$, and the ambient density 
$n_a$ (See Equation (6) \& (7)). Changes in $v_s$ and $T_s$ are relatively small 
during the H II expansion ($n_a$ = $n_0$ or 300 cm$^{-3}$), as a result we see 
the first plateau as shown in Fig. 3. The jump seen at $t$ $\sim$ 2 $\times$ 
10$^4$ yr is caused by the change from the H II expansion to the Winds phase. 
During the early Winds phase and before the shell sweeps up all the material
of its parent GMC ($t$ $<$ 0.8 Myr), the effect due to the shell deceleration 
is compensated for the effect due to the cooling in the shell. This produces
a second plateau. After this time, the shell expands into a less dense ambient
ISM, i.e. $n_a$ = $n_{ism}$ or 30 cm$^{-3}$. Less ambient pressure causes a
decrease in the shell density or a increase in the shell thickness. 
Fig.~\ref{T2Av} and Fig.~\ref{x2Av} show the profiles of the gas temperature and 
chemical abundances as a function of visual extinction $A_v$ for an expanding 
shell at several characteristic ages. The size of the PDR changes with time 
indicated by different maximum values of $A_v$ in both Fig.~\ref{T2Av} and 
Fig.~\ref{x2Av}. At $\sim$ 0.7 Myr, all mass in the GMC has been swept into 
the shell. 

\subsection{Molecular and Atomic Line Emission} \label{lrad}

The flux and intensity of FIR/sub-mm/mm line emission is calculated for several 
molecular and atomic species (CO, HCN, HCO$^+$, C, C$^+$, and O). The total 
flux or intensity of each line is the sum of the emission from the entire shell. 
For the initial 0.7 Myr, the emission from the parent GMC is also taken into 
account in the total line emission calculation. In this section, we present 
predictions of the line ratios for CO, [C I], and [C II] for an expanding shell. 
More simulations will be presented and discussed when we illustrate the model 
by a comparison with the observations of M 82 in \S~\ref{m82}.  

\begin{figure}
\epsscale{1.0}
\plotone{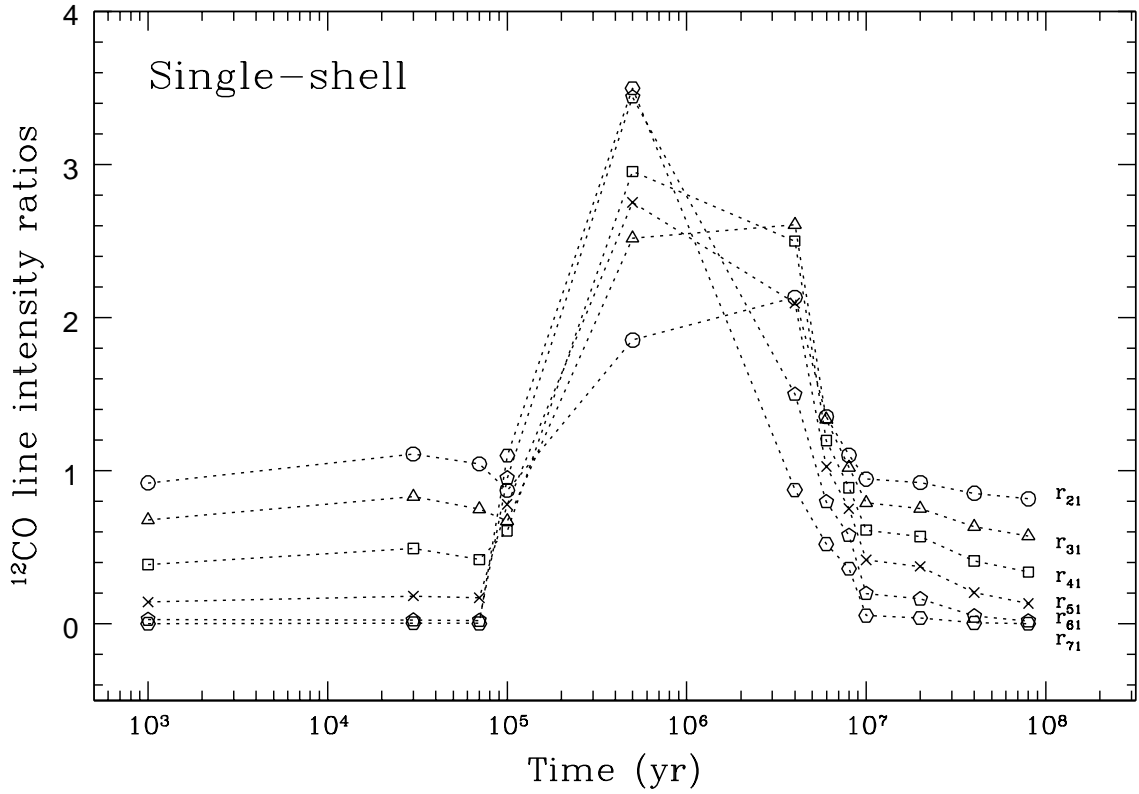}
\caption{Plot of the model $^{12}$CO line intensity ratios of high-$J$ transitions 
to the 1-0 transition as a function of starburst age for an expanding shell. 
The $^{12}$CO line intensities are compared in units of K km s$^{-1}$. \label{r2t_co}}
\end{figure}

Molecular CO is known as a good tracer for the diffuse components and total 
molecular gas content in a galaxy, but it is relatively poor tracer of the dense 
gas directly involved in massive star formation. Fig.~\ref{r2t_co} shows our 
modeling results for line ratios of high-$J$ transitions to the 1-0 transition 
of the bright and highly abundant $^{12}$CO molecule as a function of the 
starburst age of the shell. 
\begin{mathletters}
\begin{eqnarray}
r_{21} & = & I_{21} / I_{10} , \\
\vdots &  & \vdots \\
r_{71} & = & I_{76} / I_{10} 
\end{eqnarray}
\end{mathletters}
where $I_{J,J-1}$ is the line intensity, $r_{J+1,1}$ is the line intensity 
ratio, and $J$ = 1,$\ldots$,7.

For the adiabatic phase ($t$ $<$ 7.5 Myr), strong winds and supernova explosions 
compress the gas in the fast expanding shell to a high density $n$(H$_2$) $>$ 
10$^4$ cm$^{-3}$ (see Fig.~\ref{nsds}), and the strong FUV radiation $G_0$ $>$ 
10$^4$ heats up the gas and dust of the shell to a temperature $>$ 100 K 
(see Fig.~\ref{T2Av}). A significant amount of highly excited CO line emission 
is generated from the shell and its parent cloud, and therefore the line ratios 
of $r_{21}$ through $r_{71}$ are $\ge$ 1.0. At around 10 Myr, all line ratios 
(1 $\le$ $J$ $\le$ 7) have dropped below 1.0, the shell has entered the snow-plow 
phase, the corresponding FUV field $G_0$ is $\le$ 10$^4$, the shell density 
$n$(H$_2$) is $<$ 3.0 $\times$ 10$^3$ cm$^{-3}$, and the gas temperature in the 
shell $T_{gas}$ is between 20 and 230 K.

\begin{figure}
\epsscale{1.0}
\plotone{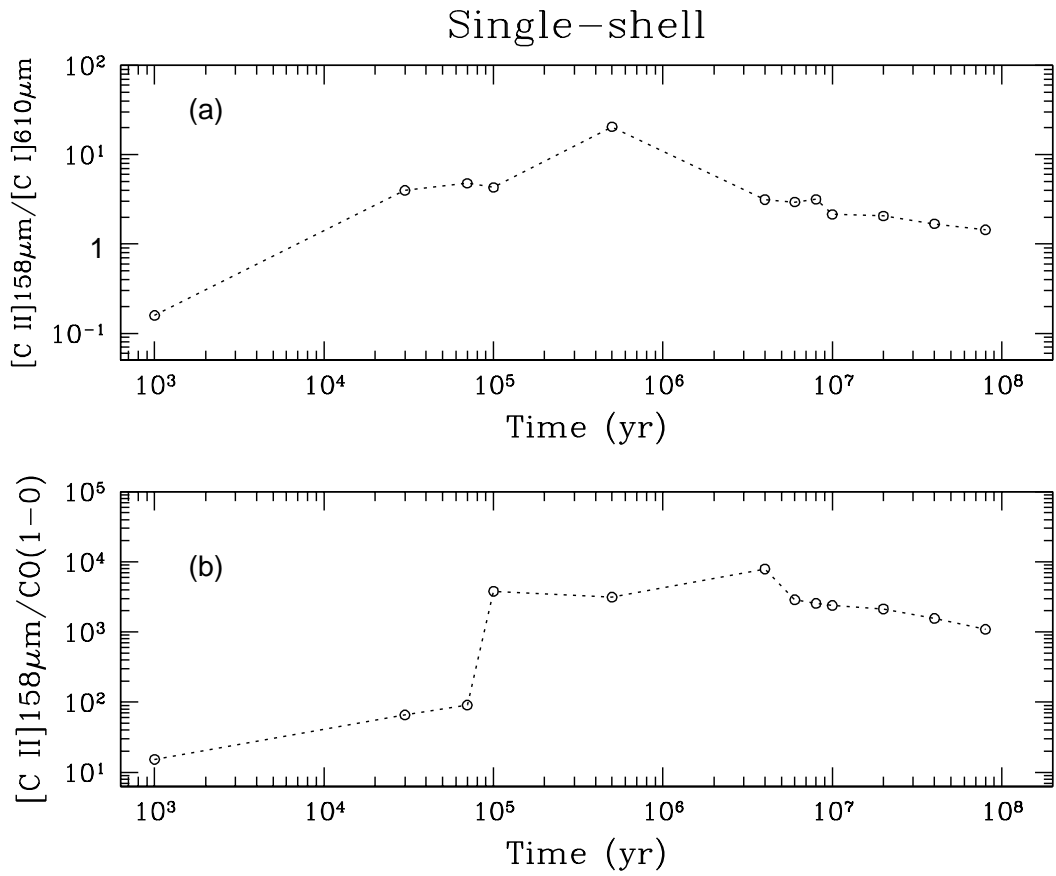}
\caption{Plots of (a) the model line intensity ratio of [C II]158$\mu$m to 
[C I]610$\mu$m as a function of time, the line intensities are compared in units 
of K km s$^{-1}$, and (b) the model line flux ratio of [C II]158$\mu$m to CO(1-0) as 
a function of time, the fluxes are compared in units of ergs cm$^{-2}$ s$^{-1}$. 
\label{r2t_atom}}
\end{figure}

The far-infrared fine structure lines are the most important cooling lines of the 
ISM in a galaxy. Fig.~\ref{r2t_atom} shows the modeling results of the line 
intensity ratio of [C II]158$\mu$m to [C I]610$\mu$m and the line flux ratio of 
[C II]158$\mu$m to CO(1-0). About 75\% of the [C II]158$\mu$m emission comes from 
PDRs and 25\% from the H II region \citep{col99}. The latter is not taken into account 
in our calculations. The [C II]158$\mu$m/CO(1-0) line flux ratio rises from 
about 10 to 10$^4$ after 1 Myr, and then slowly decreases to $\sim$ 10$^3$ 
at 80 Myr. It is clear that the cooling of the swept-up gas in the 
expanding shell is dominated by C$^+$, the contribution of the CO cooling is 
a small fraction to the total gas cooling in a massive star forming environment. 

\section{Application to The Nearby Starburst Galaxy M 82} \label{m82}

In section 4.1, we compare our modeling results with the observations of an 
expanding supershell in the nearby starburst galaxy M 82. In section 4.2, we
compare our modeling results with the average gas properties in the central 
1.0 kpc region of this galaxy. 

M 82 is an irregular starburst galaxy located at a distance of about 3.25 Mpc. 
This galaxy has been observed over a wide range of wavelengths. The starburst 
activity in M 82 was likely triggered by tidal interaction with its companion 
M 81 beginning about 10$^8$ yr ago in the nucleus, and is currently propagating
into the molecular rings. The infrared luminosity of M 82 is about 
4 $\times$ 10$^{10}$ L$_{\odot}$ arising mostly from the central $\sim$ 400 pc 
region, which has a stellar bar structure and currently has a high supernova 
rate of $\sim$ 0.05 - 0.1 yr$^{-1}$ \citep{mux94}. The evolutionary scheme in 
M 82 remains under debate. The most common suggested ages of the M 82 starburst 
in the central regions are 3 - 7 Myr predicted by Colbert et al. (1999) using one 
instantaneous burst model in dusty media with a 100 M$_{\odot}$ cut off, 
and 10 - 30 Myr predicted by Efstathiou et al. (2000) using two instantaneous 
bursts model in dusty media with a 125 M$_{\odot}$ cut off. Recently, 
F\"{o}ster-Schreiber et al. (2003) presented a more complete evolutionary scheme 
of the global starburst activity in M 82, and suggested that there are two bursts, 
one occurred at $\sim$ 5 Myr ago and another one at $\sim$ 10 Myr ago also using 
instantaneous bursts model in dusty media with a 100 M$_{\odot}$ cut off.

\subsection{The Supershell Surrounding The SNR 41.9+58} \label{superb}

Observations have detected an expanding supershell centered around the 
bright SNR 41.9+58 in both molecular line and radio continuum \citep{wei99, wil99}. 
This supershell has a diameter of $\sim$ 130 pc, an expansion velocity of 
$\sim$ 45 km s$^{-1}$, and a mass of $\sim$ 8 $\times$ 10$^6$ M$_{\odot}$. 
Using the set of initial cloud conditions selected for our simulation 
(see \S~\ref{sim}), i.e. a cloud mass $M_{GMC}$ = 10$^7$ M$_{\odot}$, 
cloud density $n_0$ = 300 cm$^{-3}$, ambient ISM density $n_{ism}$ = 30 cm$^{-3}$,
and star formation efficiency $\eta$ = 0.25, we derive a swept-up shell that 
has very similar characteristics to the observed one. At the observed radius
of $\sim$ 65 pc, our model indicates an age of 1 Myr, an expansion velocity of 
$\sim$ 45 km s$^{-1}$, and a swept up H$_2$ mass of $\sim$ 7.6 $\times$ 10$^6$ 
M$_{\odot}$.  The kinetic energy of the observed supershell is estimated
to be about 1.6 $\times$ 10$^{53}$ ergs \citep{wei99}. Our model predicts a 
kinetic energy of $\sim$ 1.5 $\times$ 10$^{53}$ ergs for the expanding shell 
at the age of 1 Myr. The total mechanical energy needed for the creation of 
this supershell is $\sim$ 1.7 $\times$ 10$^{54}$ ergs, which is contributed 
by winds and supernovae associated with $\sim$ 1700 O stars ($\ge$ 40 M$_{\odot}$). 
Therefore, about 10\% of the total energy is present in the form of kinetic 
energy of the expanding shell. The remarkably good agreement between our model 
results and the observations implies that this supershell may be created by strong 
winds and supernova explosions from a star cluster with a total mass of 
2.5 $\times$ 10$^6$ M$_{\odot}$ which occurred in the center about 1 Myr ago. 
The comparison is summarized in Table~\ref{tbl-3}. 

Furthermore, our model predictions of the CO, [C I], and [C II] line ratios for 
this expanding supershell can be used as a comparison with future observations, 
and also to constrain the physical conditions of the gas in the shell 
(see Fig.~\ref{r2t_co} and Fig.~\ref{r2t_atom} presented in \S~\ref{lrad}). 
The model line ratios which are greater than 1.0 at $t$ $<$ 8 Myr imply that 
the molecular CO is optically thin in the expanding supershell. Therefore,
it is better to look at the high CO transitions ($J$ $>$ 3) in this supershell.

\subsection{The Central Starburst Region}

Besides the known expanding supershell centered around SNR 41.9+58, there are 
other undetected shells with sizes ranging from several tens of parsec 
to more than 1 kiloparsec, and kinetic energies ranging from $\sim$ 10$^{50}$ ergs 
to more than 10$^{54}$ ergs. These shells would likely be present as partial arcs, 
or fragments, or cloud-like clumps due to strong winds and supernova explosions 
or due to shell-shell and shell-cloud interactions; only a few are visible as 
full circular arcs. The very good agreement between our model and the supershell 
observations indicates that the set of models we have put forward in this paper 
can be used to interpret other shells in a starburst galaxy like M 82. 
In this section, we illustrate the possibilities by comparing our model 
calculations with the observed FIR/sub-mm/mm properties of molecular and 
atomic line emission in the central starburst regions. 

\begin{figure}
\epsscale{1.0}
\plotone{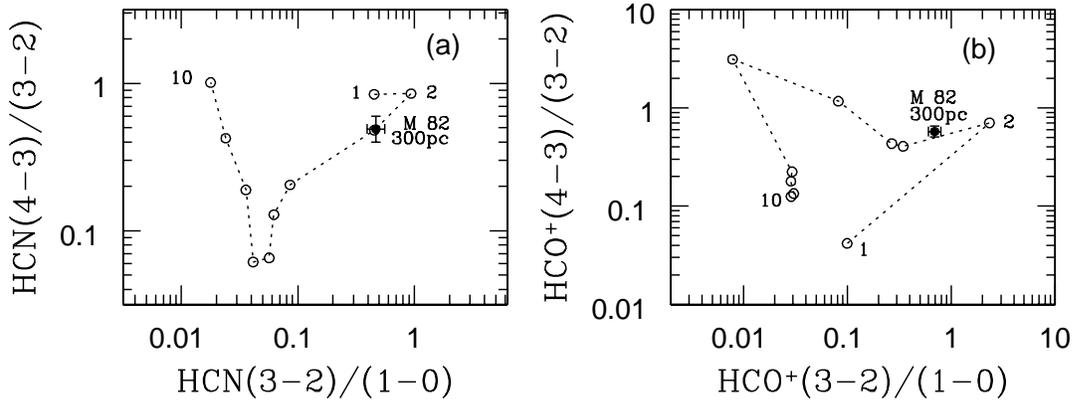}
\caption{The ratio-ratio diagrams of the HCN and HCO$^+$ line intensities. Plots of 
(a) the model HCN(4-3)/(3-2) line ratio versus the HCN(3-2)/(1-0) ratio for a sequence 
of starburst ages: 0, 0.03, 2, 6, 10, 20, 30, 40, 50, and 70 Myr 
(labeled as 1, 2,$\ldots$,10), (b) the model HCO$^+$(4-3)/(3-2) line ratio versus 
the HCO$^+$(3-2)/(1-0) ratio for a sequence of starburst ages: 0.03, 0.7, 2, 4, 10, 20, 
30, 40, 50, and 70 Myr (labeled as 1, 2$\ldots$,10). The modeling results are 
indicated by the open circles connected with dotted lines. The filled circles 
with errorbars in the plots are the observed data (see text for details). 
\label{rr2t_dens}}
\end{figure}

\begin{figure}
\epsscale{1.0}
\plotone{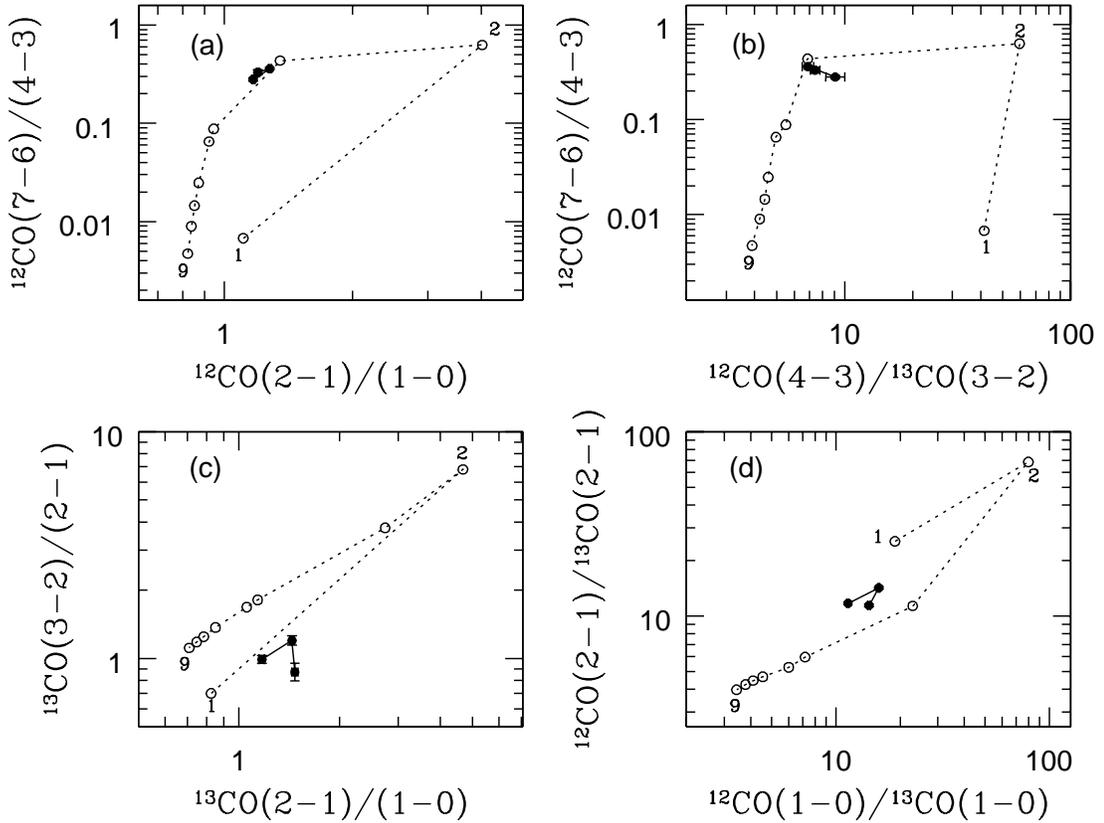}
\caption{The ratio-ratio diagrams of the $^{12}$CO and $^{13}$CO line intensities. 
The modeling results for a sequence of starburst ages: 0.03, 4, 6, 10, 20, 30, 40, 50, 
and 70 Myr (labeled as 1, 2,$\ldots$,9), are indicated by the open circles connected 
with dotted lines. The filled circles connected by solid lines show the observed data 
for the three lobes in the center of M 82 (see text for details). \label{rr2t_co}}
\end{figure}

\begin{figure}
\epsscale{1.0}
\plotone{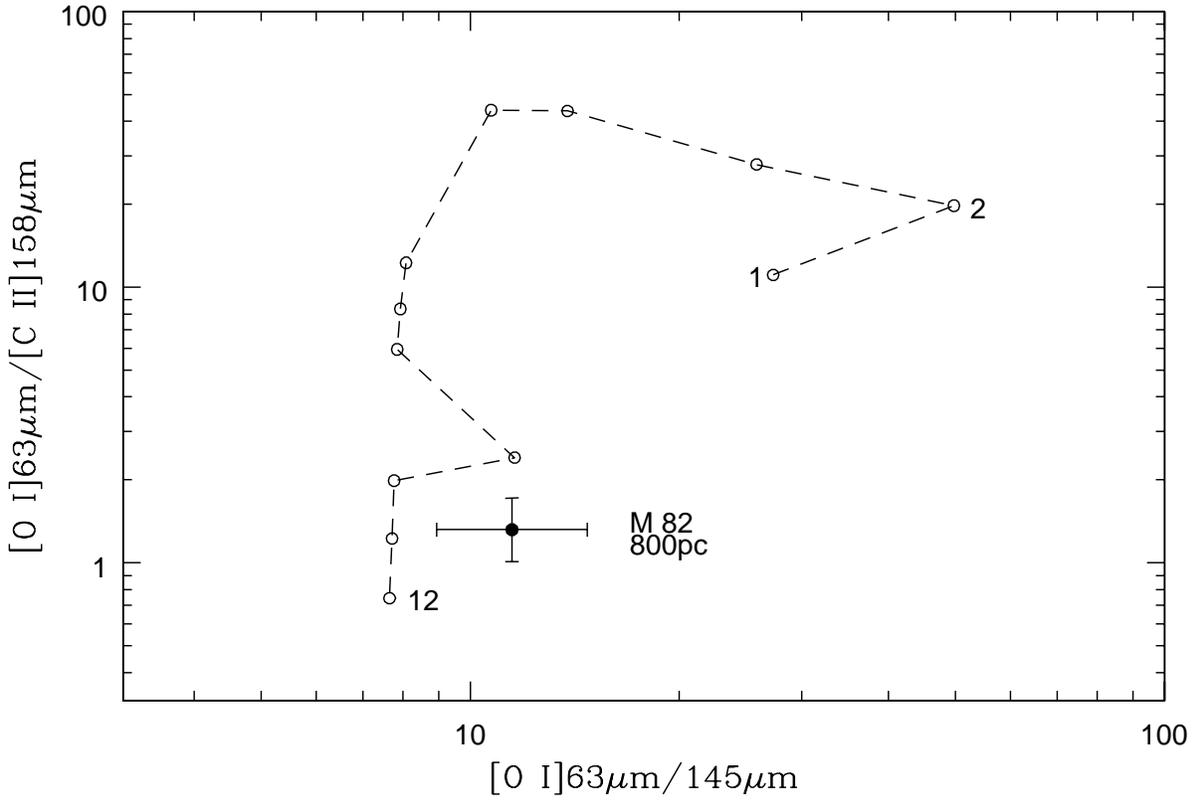}
\caption{The ratio-ratio diagram of the fine structure line fluxes. The model 
[O I]63$\mu$m/[C II]158$\mu$m ratio versus [O I]63$\mu$m/145$\mu$m ratio for 
a sequence of starburst ages: 0, 0.03, 0.07, 0.1, 0.7, 4, 6, 8, 10, 20, 40, and 80 Myr 
(labeled as 1, 2,$\ldots$,12), the line fluxes are compared in units of W m$^{-2}$. 
The modeling results are indicated by the open circles connected with dotted lines. 
The filled circles show the observed data for M 82. \label{rr2t_atom}}
\end{figure}

First of all, Fig.~\ref{rr2t_dens} shows the model ratio-ratio plots of 
HCN(4-3)/(3-2) versus HCN(3-2)/(1-0) and HCO$^+$(4-3)/(3-2) versus HCO$^+$(3-2)/(1-0), 
and a comparison with the observations of dense gas in the central 300 pc 
region \citep{saf00}. The dense gas tracers HCN and HCO$^+$ are better indicators 
of active star formation than CO, but poor tracers of the total molecular gas 
content. For both plots, the best agreement is at a starburst age of $\sim$ 3 Myr, 
implying that the expanding shell size is about 300 pc. Secondly, Fig.~\ref{rr2t_co} 
shows the model ratio-ratio plots for $^{12}$CO and $^{13}$CO, and a comparison 
with the observations of the CO line emission from three lobes (north-east, center, 
and south-west) in the central 300 to 600 pc regions \citep{mao00}. 
The isotope abundance ratio [$^{12}$CO]/[$^{13}$CO] is adopted to be 75 for the 
simulation. The best agreement shown in plots (a), (b), and (d) is at a starburst 
age of $\sim$ 6 Myr, and $\sim$ 3 Myr for plot (c), corresponding to shell sizes 
between 300 and 560 pc. The physical conditions for the gas of the shell 
at age 3 - 6 Myr are  $G_0$ $\sim$ 2 $\times$ 10$^4$ - 15 $\times$ 10$^4$, 
$n$(H$_2$) $\sim$ 1.0 $\times$ 10$^4$ - 2.0 $\times$ 10$^4$ cm$^{-3}$, 
$T_{gas}$ $\sim$ 50 - 250 K, and total molecular gas mass $M_{mol}$ $\sim$ 
0.3 $\times$ 10$^8$ - 2.1 $\times$ 10$^8$ M$_{\odot}$. Finally, Fig.~\ref{rr2t_atom} 
shows the model ratio-ratio diagram of [O I]63$\mu$m/[C II]158$\mu$m versus 
[O I]63$\mu$m/[O I]145$\mu$m, and a comparison with the observations of these 
atomic lines from the central 1.1 kpc region \citep{neg01}. The model 
[C II]158$\mu$m line is underestimated by a factor of about 1.3, since about 
25\% of the total line emission coming from the H II region is not included 
in the calculation. Therefore, the best agreement between the model and the 
observation is achieved at an age of $\sim$ 25 Myr old. The atomic line data 
are based on a 80$^{\prime\prime}$ beam whereas the molecular line data pertain 
only to the the lobes and nuclear sources at a 22$^{\prime\prime}$ beam. 
Thus, part of the reason for the discordant age in the atomic data may be the 
different regions sampled, since they may have a different starburst history. 
Our predicted gas conditions for the shell at this age are $G_0$ $\sim$ 500, 
$n$(H$_2$) $\sim$ 1.8 $\times$ 10$^3$ cm$^{-3}$, $T_{gas}$ $>$ 15 K, and 
$M_{mol}$ $\sim$ 6.0 $\times$ 10$^8$ M$_{\odot}$. These conditions are consistent
with the PDR model fits to the observations by Colbert et al. (1999). But the age 
inferred by Colbert et al. (1999) is 3 - 7 Myr. The large age discrepancy 
between the two different modeling results from the fact that our model includes 
a more massive cluster which then yields the same gas conditions (FUV flux and 
gas density) at a larger distance and hence, in the context of an expanding 
shell, an older age. It is clear that the starburst age of the whole central 
region is model dependent. More simulations with a variety of input cloud 
conditions and a comparison with data taken at a wider range of wavelengths 
are needed in order to identify the ages of starbursts accurately.  

Although different stages of development are applicable to different central 
regions of M 82, the shell sizes and the physical conditions of the gas within 
the rings (diameter $\sim$ 300 - 600 pc) predicted by our model are similar to 
what is expected from models involving expanding shells from a central starburst 
such as those proposed by Carlstrom \& Kronberg (1991). Therefore, it is possible 
that the molecular rings in M 82 are a product of gas that was swept-up by
the nuclear starburst activity which has evolved for about 10$^8$ yr. Their 
hypothesis is supported by the observations of CO line emission and continuum 
emission, as well as the discovery of supershells that have not yet had time 
to break out of the galactic plane. However, it is important to realize the 
foregoing interpretation of the lobes as a ring or torus is not unique. 
A number of authors have argued that the molecular rings may be a product of 
Linblad resonance instabilities associated with the gravitational effects of 
the bar \citep[e.g.][]{sal96, wil00}. In future work, we will carry out 
more simulations to test the hypothesis suggested by Carlstrom \& Kronberg (1991).

\section{Conclusions} \label{sum}

We have presented a set of starburst models that can be used to relate the observed 
FIR/sub-mm/mm properties of molecular and atomic gas in a starburst galaxy to 
its age and star formation history. As a preliminary approach, we have illustrated 
our model by a comparison with the observations of the expanding supershell
centered around SNR 41.9+58. The very good agreement implies that the expanding 
supershell is created by strong stellar winds and SN explosions from a young 
star cluster ($\sim$ 2.5 $\times$ 10$^6$ M$_{\odot}$) located in its center. 

Our model predictions of CO, HCN, and HCO$^+$ line ratios agree with the molecular 
data for the central lobes (300 - 600 pc) for a shell with an age in the 3 - 7 Myr 
range. This implies that the molecular rings are possibly a consequence of 
swept-up or compressed gas caused by massive star formation originating in the 
nucleus of M 82. More simulations in future work may be able to justify
this hypothesis. The atomic line ratios calculated by our model do not fit the
observed data as well as the molecular data, but suggest a much older shell, 
because the atomic line emission comes from a much larger region ($>$ 1 kpc). 
A variety of modeling parameters need to be considered to yield more accurate
starburst ages.

Our model also yields appropriate values for the gas density, temperature, 
and structure scales compared to those measured in the center of M 82 
\citep[e.g.][]{las63, rie89, stu97, saf00, mao00, neg01, war03}, and the 
total H$_2$ content within the inner 600 pc ($\sim$ 2.0 $\times$ 10$^8$ M$_{\odot}$) 
is compatible with the observations \citep[e.g.][]{wil92}. Therefore, 
the neutral ISM in the central star-forming region of M 82 may be viewed as a 
product of evolving shells, and is now presenting itself in the form of 
fragments, small cloud clumps, sheets, or even full circular arcs.

\acknowledgments

LY would like to thank Dr. Andreas Efstathiou, Dr. Claus Leitherer, and 
Dr. Peter van Hoof. LY is grateful to Dr. Chris Loken and Mr. Hugh Merz at 
the Canadian Institute for Theoretical Astrophysics for their generosity in 
allowing her to run the PDR simulations on their fastest PC machines. LY is 
grateful for use of the PPARC funded Miracle Computing Facility, located at 
UCL, to perform her RT calculations. We thank the referee for helpful 
comments and suggests. This research was supported by a research grant from the 
Natural Sciences and Engineering Research Council of Canada to ERS, and a 
Reinhardt Graduate Student Travel Fellowship from the Department of Astronomy 
and Astrophysics at the University of Toronto. TAB is supported by a PPARC 
studentship. SV acknowledges individual financial support from a PPARC advanced 
Fellowship.



\clearpage

\clearpage

\clearpage

\clearpage

\clearpage

\clearpage

\clearpage

\clearpage

\clearpage

\clearpage

\clearpage

\begin{deluxetable}{ll}
\rotate
\tablecaption{Summary of Starburst Models For FIR/sub-mm/mm Line Emission. \label{tbl-1}}
\tablewidth{0pt}
\tablehead{
\colhead{Name} & \colhead{Description}   
}
\startdata
Assumptions:   & - spherical geometry, non-magnetized (GMCs and shells)   \\
               & - no interactions between shells, or shell and cloud \\
               & - dustless H II regions \\
               & - uniform densities of GMCs and ambient media \\
               & - all stars form instantaneously, no stars form inside the shells \\
               & - stellar mass 0.1 - 120 M$_{\odot}$ with Salpeter IMF 
                   $dN/dM_{\ast}$ $\propto$ $M^{-2.35}_{\ast}$ \\
               & - PDRs exist primarily within the expanding shells \\ \hline    
Input Parameters: & - GMC mass $M_{GMC}$ = 10$^7$ M$_{\odot}$ \\
               & - average cloud density $n_0$ = 300 cm$^{-3}$, cloud core density 
                  $n_c$ = 2000 cm$^{-3}$ \\
               & - ambient ISM density $n_{ism}$ = 30 cm$^{-3}$ \\
               & - star formation efficiency $\eta$ = 0.25 \\ 
               & - metallicity $\mathcal{Z}$ = 1.0 Z$_{\odot}$ \\ 
               & - gas-to-dust ratio = 100 \\ \hline
Output Parameters: & - radius, velocity, temperature, density, and thickness of the 
                       shell \\
               & - chemical abundances of different molecules and atoms in the shell \\
               & - integrated line intensity/flux, line ratios \\ \hline 
Observational  & - line intensities/fluxes and line ratios for molecules 
                   (e.g. $^{12}$CO, $^{13}$CO, HCN, HCO$^+$), \\
               & $ $ $ $ and atoms (e.g. [C I], [C II], [O I]) \\ \hline
\enddata
\tablecomments{See also Table~\ref{tbl-2} for more input parameters for the 
time-dependent PDR model.}
\end{deluxetable}

\clearpage 

\begin{deluxetable}{lll}
\tablecaption{Input Parameters For The Time-dependent PDR Model. \label{tbl-2}}
\tablewidth{0pt}
\tablehead{
\colhead{Parameter} & \colhead{Symbol} & \colhead{Value} 
}
\startdata
Starburst age (yr) & $t$ & 0 $\le$ $t$ $<$ 10$^8$ \\
Incident FUV flux (Habing field) & $G_0$ & 10 $<$ $G_0$ $\le$ 10$^8$ \\ 
Turbulent (microturbulence) velocity (km s$^{-1}$) & $\delta$$v_D$ & 1.5 \\
PDR surface density ($A_v$ = 0 mag) & $n_H$ &  10$^3$ $\le$ $n_H$ $<$ 10$^7$ \\
Initial gas-phase abundances relative to H\tablenotemark{a}  &  & \\
PAH abundance & $x_{PAH}$ & 4.0 $\times$ 10$^{-7}$ \\
Dust visual absorption cross section (cm$^{-2}$) & $\sigma_v$ & 3.1 $\times$ 10$^{-10}$ \\
H$_2$ formation rate on dust at $A_v$ = 0 (cm$^3$ s$^{-1}$) & $\eta_{H_2}$ & 3.0 $\times$ 10$^{-18}$ \\ 
Cosmic-rays ionization rate (s$^{-1}$) & $\zeta$ & 1.3 $\times$ 10$^{-17}$ \\ \hline                   
\enddata
\tablenotetext{a}{The initial gas-phase abundances for all depths at the first 
time-step ($t$ = 0 yr) are produced by a single-point dense dark-cloud model 
(see text for details).}
\end{deluxetable}

\clearpage

\begin{deluxetable}{lll}
\tablecaption{Characteristics of The Expanding Supershell in M 82. \label{tbl-3}}
\tablewidth{0pt}
\tablehead{
\colhead{Parameter} & \colhead{Observation} & \colhead{Model} 
}
\startdata
Radius (pc) & 65.0 & 65.0 \\
Age (Myr)   & 1.0  &  1.0 \\
Expansion velocity (km s$^{-1}$) & 45 & 45 \\ 
Total H$_2$ molecular gas mass ($\times$ 10$^6$ M$_{\odot}$) & 8.0 & 7.6 \\
Kinetic Energy ($\times$ 10$^{53}$ ergs) & 1.6 & 1.5 \\
Total stellar mass in the center cluster ($\times$ 10$^6$ M$_{\odot}$) & $\ldots$ & 2.5  \\
Total number of O stars ($\ge$ 40 M$_{\odot}$) & $\ldots$   & 1700 \\
Total Mechanical Energy ($\times$ 10$^{54}$ ergs) & $\ldots$ & 1.7  \\ \hline            
\enddata
\end{deluxetable}

\end{document}